\documentclass[aps,prd,reprint,twocolumn,superscriptaddress,nofootinbib,floatfix,showpacs]{revtex4-1}
\pdfoutput=1
\usepackage[T1]{fontenc}
\usepackage[utf8]{inputenc}
\usepackage{amsmath,amssymb,amsfonts}
\usepackage{mathrsfs}
\usepackage{bbm}
\usepackage{slashed}
\usepackage[dvipdf,dvips,dvipdfmx]{graphicx}
\usepackage{verbatim}
\usepackage[english]{babel}
\usepackage{mathbbol}

\def\apjl{ApJ}%
\def\apjs{ApJS}%
\def\ao{Appl.\ Opt.}%
\def\aap{A\&A}%
\def\jcap{JCAP}%
\def\prd{Phys.\ Rev.\ D}%
\def\physrep{Phys.\ Rep.}%
\usepackage{aas_macros}
\usepackage[colorlinks,citecolor=blue,linkcolor=blue,urlcolor=blue]{hyperref}

\def\be{\begin{equation}}
\def\ee{\end{equation}}
\def\ba{\arraycolsep .1em \begin{eqnarray}}
\def\ea{\end{eqnarray}}

\begin{document}

\title{Dynamical renormalization of black-hole spacetimes} 

\author{Alessia Platania}
\affiliation{Institut für Theoretische Physik, Universität Heidelberg, Philosophenweg 16, 69120 Heidelberg, Germany}

\begin{abstract}
We construct a black-hole spacetime which includes the running of the gravitational coupling in a self-consistent way. Starting from a classical Schwarzschild black hole, the backreaction effects produced by the running Newton's coupling are taken into account iteratively. The sequence, described by a simple recurrence relation, flows towards a self-consistent solution that can be derived analytically. As a key result, if the gravitational renormalization group flow attains a non-trivial fixed point at high energies, the sequence converges to a ``renormalized'' black-hole spacetime of the Dymnikova-type, which is free of singularities.
\end{abstract}

\maketitle

\section{Introduction} \label{sect1}

The Wilsonian formulation of the Renormalization Group (RG) has undoubtedly led to a deeper understanding of the concept of renormalization in quantum field theory. 
One the most compelling consequences of the Wilsonian idea of renormalization is the possibility of constructing a well-defined quantum theory for the gravitational interaction within the standard and well-tested framework of quantum field theory. In fact, although General Relativity is perturbatively non-renormalizable, the existence of a non-trivial fixed point \citep{Wilson:1973jj} of the gravitational RG flow might allow for a safe quantization of the gravitational field. This idea, first conjectured by Weinberg \cite{1976W,1979W}, paved the way for the so-called asymptotic safety scenario for quantum gravity. Starting from the formulation of a flow equation for the gravitational interaction \cite{martin} based on the modern functional renormalization group (FRG) techniques \cite{Benedetti:2010nr}, indications for the existence of a fixed point underlying the non-perturbative renormalizability of gravity have been found in a number of different approximation schemes \cite{Souma:1999at,Lauscher:2001ya,Reuter:2001ag,Lauscher:2002sq,Codello:2006in,Machado:2007ea,Benedetti:2009rx,Groh:2011vn,Dietz:2012ic,Benedetti:2013jk,Eichhorn:2013xr,Falls:2014tra,Ohta:2015efa,Eichhorn:2015bna,Gies:2016con,Henz:2016aoh,Denz:2016qks,Alkofer:2018fxj}, in the case of spacetimes carrying a foliation structure \cite{Manrique:2011jc,Rechenberger:2012dt,Biemans:2016rvp,Biemans:2017zca,Houthoff:2017oam,Knorr:2018fdu}, and also in the presence of matter fields \cite{Percacci:2003jz,Vacca:2010mj,Dona:2012am,Dona:2013qba,Meibohm:2015twa,Dona:2015tnf,Christiansen:2017cxa}.

The gravitational quantum effective action is generated by a shell-by-shell integration of the fluctuating modes of the quantum gravitational field. 
Accordingly, quantum fluctuations modify the effective dynamics of the theory: the strength of its interaction depends on the energy scale $k$ at which a phenomenon is observed. Provided that the asymptotic safety condition is realized, at high energies the Newton's coupling scales as~$G_k\sim g_\ast k^{-2}$, $g_\ast$ being the UV-value of the dimensionless Newton's coupling. The scale-dependent gravitational coupling $G_k$ thus vanishes at short-distances. Due to this anti-screening behavior, one would expect substantial modifications to the microscopic structure of the effective spacetime geometry, that might even acquire fractal properties \cite{Lauscher:2005qz,Reuter:2011ah,Calcagni:2013vsa}.

Since the RG equations describe the running of the action with the energy scale, introducing the running of the couplings in the solutions to a classical field theory allows for a semi-classical description of the fully quantized theory and a qualitative understanding of its dynamics \cite{Polonyi:2001se}. Employing this strategy, commonly referenced as RG-improvement procedure \cite{cw,1973migdal,1973gross,1978pagels,1978Matinyan,1983adler,Dittrich:1985yb}, the implications of an anti-screening gravitational coupling have been broadly investigated. As a key finding, a UV-vanishing dimensionful Newton's coupling might entail an effective weakening or even avoidance of the classical singularities typically appearing in cosmological contexts \cite{Kofinas:2015sna,Kofinas:2016lcz,Bonanno:2017gji} and in the black-hole interior \cite{br00,2006alfiomartin,Falls:2010he,Koch:2013owa,Bonanno:2016dyv,Torres:2017ygl,Pawlowski:2018swz,Adeifeoba:2018ydh}. In particular, as pointed out in~\cite{Adeifeoba:2018ydh}, the divergence of the classical Kretschmann scalar inside a Schwarzschild black-hole might be a consequence of the negative mass dimension of the Newton's constant. 

A technical and conceptual issue in the implementation of the RG-improvement procedure is to understand how to construct the energy scale $k$ relates to the typical length scale of the phenomena under consideration. 
While the answer to this question seems obvious in flat spacetimes, where $k\sim l^{-1}$, the same does not hold when gravity is turned on.
An explicit construction of the mass scale $k$ within semi-classical approaches makes manifest use of a classical background metric, which is no longer valid (and even singular) in the strong-curvature regime. In addition, introducing the scale-dependence of the couplings at the level of the classical solutions, entails a modification to the Einstein field equations. This modification can be encoded in an effective energy-momentum tensor \cite{Reuter:2004nx,Koch:2010nn}, encoding vacuum polarization effects of the quantum spacetime geometry \cite{Poisson:1988wc,Poisson:1990eh}. This backreaction, in turn, modifies the effective spacetime metric and the cutoff function itself. 

The focus of this paper is the construction of an effective spherically-symmetric black-hole spacetime, where the running of the gravitational coupling is incorporated self-consistently by means of an iterative procedure. The iteration relies on a self-adjusting cutoff, that is independent of the initial background metric, and naturally accounts for the backreaction produced at the level of the (modified) Einstein field equations. In addition, the sequence of RG-improvements is described by a simple recurrence relation, whose fixed-point solution can be derived analytically. As it will be shown, the running of the gravitational coupling triggered by quantum fluctuations induces an ``effective renormalization'' of the bare spacetime geometry, which dynamically removes the singularity affecting the initial classical metric. Specifically, starting from a (classical) Schwarzschild black-hole, the sequence of RG-improvements flows towards a self-consistent solution that, under the condition of non-perturbative renormalizability of gravity, is a (regular) Dymnikova black-hole \cite{Dymnikova:1992ux,Dymnikova:1996ux}. The running of the Newton's constant in fact generates an effective tension which opposes and perfectly counterbalances the action of the classical tidal forces, thus leading to singularity-resolution. 

This Letter is organized as follows: Section~\ref{sect2} discusses in detail the problem and introduces the basic formalism. In Section~\ref{sect3} we setup an iterative procedure aimed at finding an effective black-hole spacetime based on a self-adjusting cutoff. The resulting self-consistent solution is derived and discussed. Finally, Section~\ref{sect4} is devoted to a summary and discussion of the results obtained. 

\section{Discussion of the problem and basic setup} \label{sect2}

The Renormalization Group (RG) provides a tool to investigate the effective dynamics of a system at any arbitrary energy scale~$k$. This scale sets the ``resolution'' of the \textit{RG-microscope}. Fluctuating modes with momenta $p \gtrsim k$ are integrated out, resulting in a scale-dependent effective action~$\Gamma_k$ describing the theory in terms of effective interactions between fundamental ``blocks'' of linear dimension~$\sim k^{-1}$. In the continuum limit $k\to\infty$, the action functional~$\Gamma_k$ gives the bare action of the theory, whilst for $k\to0$ it reproduces the standard full effective action. The flow of $\Gamma_k$ between these two limits is described by the Functional Renormalization Group Equation (FRGE) \cite{Wetterich:1992yh,Morris:1993qb}
\be \label{floweq}
k\partial_k \Gamma_k=\frac{1}{2}\mathrm{STr}\left\{(\Gamma_k^{(2)}+\mathcal{R}_k)^{-1} k\partial_k \mathcal{R}_k \right\}\;,
\ee
$\mathcal{R}_k$ being an infrared regulator. It acts as an effective mass-squared term, $\mathcal{R}_k\sim k^2$, that decouples the modes with $p \gtrsim k$, from low-energy quantum fluctuations. The insertion of this regulator into the action at the level of the functional integral is the key ingredient to perform a shell-by-shell integration of the fast-fluctuating modes and therefore to generate the flow of the scale-dependent functional $\Gamma_k$.

The modifications to the spacetime geometry induced by quantum effects are to be found by solving the quantum equations of motion generated by the scale-dependent action $\Gamma_k$ \cite{Lauscher:2005qz}
\be \label{EOM}
\frac{\delta \Gamma_k}{\delta g_{\mu\nu}}[\langle g_{\mu\nu} \rangle_k]=0 \;.
\ee
It is important to stress that a given solution~$\langle g_{\mu\nu}\rangle_k$ is a (semi-)classical field describing how the quantum-fluctuating spacetime $g_{\mu\nu}$ looks like at the energy scale~$k$: changing~$k$ is equivalent to changing the level of detail of our description of the spacetime under the RG-microscope \cite{Lauscher:2005qz}.

Starting from a classical action, eq.~\eqref{floweq} provides the running of the couplings with respect to the RG cutoff $k$. This running qualitatively agrees with the actual variation of the full quantum theory under changes of the energy scale \cite{Polonyi:2001se} and it can be used as a tool to understand what is the impact of quantum fluctuations on the classical solutions of a field theory. In fact, since the variation of the effective interactions in $\Gamma_k$ is induced by quantum fluctuations, introducing the running of the couplings in the classical solutions of a field theory is expected to effectively mimic quantum effects, and might give important hints on how the classical solutions are modified at high energies. This is the basic idea underlying the so-called RG-improvement procedure: the coupling constants of a classical field theory are replaced by the corresponding running couplings, giving rise to an effective description of the quantum theory as a function of the RG scale~$k$. Subsequently, the energy scale $k$ is identified with the typical momentum scale of the phenomenon under consideration, e.g. the center-of-mass energy of a scattering process \cite{Litim:2007iu,Gerwick:2011jw}, or the corresponding characteristic length. For field theories on a fixed flat background, this typical length scale is $l\sim k^{-1}$. This semi-classical method has been widely used in the context of Quantum Field Theory \cite{cw,1973migdal,1973gross,1978pagels,1978Matinyan,1983adler} and in simple cases provides a good approximation to the actual quantum dynamics \cite{Dittrich:1985yb,2011eichhorngluon}. 

Here we want to apply this strategy to investigate the high-energy modifications to the spacetime geometry of a classical black hole. As will be soon apparent, the generalization of the method described above to the case of gravity is not straightforward. For simplicity, we specialize to the case of a Schwarzschild spacetime in ingoing Eddington–Finkelstein coordinates. The line element reads
\be \label{metric}
ds^2=\bar{g}_{\mu\nu}dx^\mu dx^\nu=-f(r)\,dv^2+2drdv+r^2d\Omega^2\;,
\ee
where $\bar{g}_{\mu\nu}$ is the classical metric and the lapse function~$f(r)$ is given by
\be \label{lapse}
f(r)=1-\frac{2mG_0}{r} \;,
\ee
$m$ being the black-hole mass, and~$G_0$ being the Newton's constant measured at infinity. The classical curvature invariants, the Ricci and Kretschmann scalars, read
\be \label{classcalars}
\bar{R}_\mathrm{cl}=0\;,\qquad\bar{K}_\mathrm{cl}=\bar{R}_{\mu\nu\sigma\rho}\bar{R}^{\mu\nu\sigma\rho}=\frac{48 m^2 G_0^2}{r^6}\;,
\ee
Due to the divergence of the tidal forces $\bar{K}_\mathrm{cl}$, the classical spacetime develops a singularity at~$r=0$.

The typical momentum scale $k$ of the scattering processes occurring in a black-hole spacetime depends on the distance from the classical singularity. The effective description~\eqref{EOM} of the spacetime dynamics as seen by an observer moving along a radial geodesic, $x_\mu(\tau)$, and approaching the black-hole, thus depends on the spacetime point, $k=k[x_\mu(\tau)]$. 
In a first approximation, the metric perceived by the observer is described by the line element~\eqref{metric}, with the Newton's constant replaced by its running counterpart $G_k$
\be \label{Newton}
G_k=\frac{G_0}{1+g_\ast^{-1}\,G_0 k^2} \;\;.
\ee
This scaling relation was derived in \cite{br00} by solving the FRG equations~\eqref{floweq} in the Einstein-Hilbert approximation, with cosmological constant set to zero. Here $k=0$ corresponds to the classical regime, where $G_k$ is constant. In the opposite limit $k\gg M_{Pl}$, the Newton's coupling scales as $G_k\sim g_\ast k^{-2}$, $g_\ast$ being the fixed-point value of the dimensionless Newton's coupling.

The computation of the quantum-improved metric requires a scale identification $k=k(r)$ which associates the departure from ``classicality'' of the spacetime with the dynamics of the gravitational RG flow: the spacetime is classical at low energies, $k\ll M_{Pl}$, while quantum corrections are expected to play an important role at Planckian energies, i.e. in the proximity of the classical singular hypersurface $r=0$. It is reasonable to assume that $k(r)$ is a monotonically decreasing function of the radial coordinate $r$ (by symmetry reasons, $k$ cannot depend on other coordinates). In particular, a natural generalization of the scale-setting $k\sim r^{-1}$ employed in the case of flat spacetimes \cite{Dittrich:1985yb}, might be $k\sim 1/d_\mathrm{class}(r)$ \cite{br00,Koch:2013owa}, where $d_\mathrm{class}(r)$ is the classical proper distance
\be \label{classprop}
d_\mathrm{class}(r)\sim \frac{r^{3/2}}{\sqrt{G_0\,m}}\;.
\ee
Using this identification, in \cite{br00} a regular RG-improved black-hole solution, corresponding to the well-known Hayward black hole \cite{Hayward:2005gi}, was derived for the first time. 

The classical proper distance $d_\mathrm{class}(r)$, or any other scalar quantity providing an effective infrared-cutoff scale~\cite{Bonanno:2016dyv}, is constructed using the classical background metric. Accordingly, this construction runs into several interrelated problems:
\begin{itemize}
\item The cutoff scale $k^2=\bar{g}_{\mu\nu}k^\mu k^\nu$ manifestly depends on the initial classical background $\bar{g}_{\mu\nu}$, obtained by solving the classical field equations. In the strong-curvature regions, $k(x)\gg M_{Pl}$, General Relativity is expected to break down and $\bar{g}_{\mu\nu}$ cannot give an accurate description of the spacetime anymore. This means that a proper construction of $k$ would require the knowledge of the effective metric $\langle g_{\mu\nu}\rangle_k$ itself \cite{Lauscher:2005qz}, which is a priori unknown.
\item The classical spacetime metric $\bar{g}_{\mu\nu}$ explicitly depends on the bare Newton's constant $G_0$. As a consequence, when quantum-corrections are taken into account within semi-classical approaches, the metric is modified by the running of the Newton's coupling and the scale setting $k=k(r)$, e.g. eq.~\eqref{classprop}, used to derive the RG-improved metric does not hold anymore. 
\item Forcing the original black-hole spacetime to be compatible with the running of the Newton's coupling induces a modification of the spacetime dynamics at the level of the Einstein equation \cite{Reuter:2004nx,Koch:2010nn}. This backreaction effect has to be taken into account.
\end{itemize}
All these points make it clear that the application of the RG-improvement procedure to gravity requires to construct simultaneously an effective  metric $\langle g_{\mu\nu}\rangle_{k(x)}$ and a cutoff $k^2=\langle{g}_{\mu\nu}\rangle_k k^\mu k^\nu$ built by means of $\langle g_{\mu\nu} \rangle_k$ itself.

Starting from the classical metric \eqref{metric}, the aim of the present work is to implement an iterative procedure that, relying on a self-adjusting cutoff, takes into account the backreaction effects produced by the variation of the gravitational coupling. This will allow us to construct a self-consistent RG-improved black-hole metric, compatible with a running Newton's coupling of the form~\eqref{Newton}.

\section{Iterative RG-improvement procedure} \label{sect3}

Let us start from the classical Schwarzschild geometry. The structure of the metric is given by the line element~\eqref{metric} and corresponds to a solution to the \textit{vacuum} Einstein field equation with lapse function \eqref{lapse}
\be \label{startsys}
\begin{cases}
R_{\mu\nu}-\frac{1}{2}R\,g_{\mu\nu}=0\\[0.2cm]
f_{(0)}(r)=1-\frac{2\,m\,G_{0}}{r}
\end{cases}
\ee
We now perturb the system \eqref{startsys} by replacing the Newton's constant $G_0$ with its running counterpart
\be \label{firstrepl}
G_{0}\to G(r)=\frac{G_0}{1+g_\ast^{-1} G_0 \, k^2(r)}\;,
\ee
$k(r)$ being a cutoff function such that $k(r)$ goes to zero as $r\to\infty$. The latter condition guarantees that the classical Schwarzschild metric re-emerges in the classical region, i.e. $\bar{g}^{\mu\nu}\equiv\langle g^{\mu\nu}\rangle_k$ for $k\ll M_{Pl}$ and $r\gg l_{Pl}$. The replacement~\eqref{firstrepl} gives rise to a new metric of the form~\eqref{metric}, with lapse function
\be
f(r)=1-\frac{2mG[k(r)]}{r}\;.
\ee
This modified spacetime can be thought of as an exact solution of the Einstein equations in the presence of an \textit{effective} energy-momentum tensor of the form \cite{1999WW}
\be \label{Teff}
T_{\mu\nu}^{\text{eff}}=(\rho+p)(l_\mu n_\nu+l_\nu n_\mu)+p g_{\mu\nu}\;.
\ee
Here $l_\mu$ and $n_\mu$ are null vectors satisfying $l_\mu n^\mu=-1$, while the energy density $\rho$ and pressure $p$ are generated by the variation of the Newton's constant with the radial coordinate $r$ \cite{1999WW}
\be \label{qgfluid}
\rho=\frac{m\,G'(r)}{4\pi r^2\,G(r)}\;\;,\qquad p=-\frac{m\,G''(r)}{8\pi r\,G(r)}\;.
\ee
The effective energy-momentum tensor~\eqref{Teff} should be understood as a result of the vacuum polarization effects of the quantum gravitational field~\cite{Poisson:1988wc,Poisson:1990eh}. The corresponding energy-density~$\rho$ can then be interpreted as an effective \textit{quantum-gravitational self-energy}. The quantum system is self-sustaining: a small variation of the Newton's constant triggers a ripple effect, consisting of successive back-reactions of the semi-classical background spacetime which, in turn, provokes further variations of the Newton's coupling. Based on these arguments, the energy-density~$\rho\propto\partial_r \mathrm{log}\,G(r)$ might give a measure of the strength of quantum effects inside the horizon, and it can be used to self-consistently construct the cutoff function~$k_{(n+1)}(r)$ for $n>1$. The RG-improvement iteration is then formalized as follows.

The step ``0'' of the iteration corresponds to the classical Schwarzschild spacetime and is characterized by the parameters
\be
k_{(0)}(r)=0 \quad\Rightarrow\quad G_{(0)}=G_0\;\;,\;\; T_{\mu\nu}^{\mathrm{eff}(0)}=0\;.
\ee
The first step is defined by the replacement~\eqref{firstrepl}, where $k=k_1(r)$ is chosen arbitrarily and serves as an initial condition to perturb the system \eqref{startsys}. Successive steps of the iterative procedure, $n>1$, are defined by the replacement
\be
G_{(n)}\to G_{(n+1)}(r)=\frac{G_0}{1+g_\ast^{-1} G_0 \, k_{(n+1)}^2(r)}\;\;,
\ee
where the cutoff function $k_{(n+1)}(r)$ is constructed as a functional of the energy-density $\rho_{(n)}(r)$ generated by the variation of $G(r)$ in the previous step
\be \label{selfcutoff}
k_{(n+1)}^2(r)\equiv\mathcal{K}[\rho_{(n)}(r)]\;.
\ee
The procedure generates a sequence of modified Einstein equations, admitting solutions of the form \eqref{metric} with lapse functions $f_{(n)}(r)$
\be \label{RGiteration}
\begin{cases}
R_{\mu\nu}-\frac{1}{2}R\,g_{\mu\nu}= G_{(n+1)}(r) \,\,T_{\mu\nu}^{\text{eff}\,(n+1)}\\[0.2cm]
f_{(n+1)}(r)=1-\frac{2\,m\,G_{(n+1)}(r)}{r} \;\;.
\end{cases}
\ee
The running Newton's coupling at the step~$(n+1)$ depends on the \textit{self-adjusting} cutoff $k_{(n+1)}$ which, in turn, is determined by the self-energy density generated by the variation of the Newton's coupling at the step~$n$, $\rho_{(n)}\propto \partial_r G_{(n)}$. As a result, the sequence of RG-improvements~\eqref{RGiteration} is completely determined by the following recursive relation
\be \label{Gnsequence}
G_{(n+1)}(r)=\frac{G_0}{1+g_\ast^{-1} G_0 \, \mathcal{K}[G'_{(n)}(r)]}\;.
\ee
Once the functional form of $\mathcal{K}$ is specified, and assuming that the sequence of improvements defined in eq.~\eqref{RGiteration} converges, a fixed point can be found by considering the limit $n\to\infty$. If this limit exists, the sequence \eqref{RGiteration} tends to a metric of the form \eqref{metric}, with a scale-dependent Newton's constant $G(r)\equiv G_{(\infty)}(r)$ defined by
\be \label{fpeq}
\mathcal{K}[G'_{(\infty)}(r)]=g_\ast \frac{G_0-G_{(\infty)}(r)}{G_0 \, G_{(\infty)}}\;.
\ee
At this point we only need to specify the $\mathcal{K}$-functional in eq.~\eqref{selfcutoff}. In the case of the Einstein-Hilbert truncation with non-vanishing cosmological constant, there exists a precise relation between the cutoff $k$ and the energy density appearing in the right-hand-side of the Einstein field equation. This relation arises from the requirement to preserve the general covariance of the theory \cite{guberina05}. In the case at hand, imposing the Bianchi identity does not give information on the relation~\eqref{selfcutoff} and physical arguments are required in order to understand what is the physical infrared cutoff $k(r)$. To this end it is crucial to see how the classical Ricci and Kretschmann scalars \eqref{classcalars} are modified by the variation of Newton's coupling.

Approaching the classical singularity, the classical tidal forces increase as~$\sim1/r^6$ and diverge at $r=0$. Quantum-gravity effects are expected to produce an opposite tension which overcomes them, and regularizes the spacetime by weakening the resulting quantum-corrected tidal forces. The energy-density~$\rho$ might produce a similar effect: inside the horizon time and space exchange their roles, and the effective self-energy-density~$\rho$ produced by the variation of the Newton's coupling acts like a tension along the normal to the hypersurfaces $r=\text{const}$ \cite{Poisson:1988wc,Poisson:1990eh}. The presence of the effective energy-momentum tensor in eq.~\eqref{Teff} thus results in a balance between the classical tidal forces and their quantum counterpart. This is explicitly seen by analyzing the quantum-corrected Kretschmann scalar
\begin{align} \label{QKre}
K_{(n)}=&\frac{G_{(n)}^2}{G_0^2}  \left\{ \bar{K}_\mathrm{cl} -\frac{32\pi(2+\omega)}{\sqrt{3}}{\sqrt{\bar{K}_\mathrm{cl}}}\,{\left(\rho_{(n)}G_0 \right)}\right. \nonumber \\ 
& \left.\;\;+256\pi^2(2+2\omega+\omega^2)\,{\left( \rho_{(n)}G_0 \right)^2 }\right\}\;,
\end{align}
where $w=w_{(n)}(r)$ is a dimensionless function, defining the equation of state $p=\omega \rho$ of the effective fluid~\eqref{qgfluid}.
The strength of the classical tidal forces, described by~$\bar{K}_\mathrm{cl}$, is now modulated by the pre-factor~$G^2(r)/G_0^2$. At the same time, $\bar{K}_\mathrm{cl}$ is counterbalanced by two additional terms whose magnitude depends on the mass scale $(G_0\,\rho)$. Similarly, the tension generated by $\rho$ induces a non-vanishing curvature
\be \label{QRic}
R_{(n)}=\frac{G_{(n)}}{G_0}\left\{\bar{R}_\mathrm{cl}+16\pi(1-w)\left( \rho_{(n)}G_0 \right) \right\}
\ee
The characteristic mass scale $(G_0\,\rho)$ emerges naturally in the expressions of the quantum-improved Ricci and Kretschmann scalars, and could technically act as a scale-dependent regulator for the bare curvature invariants. Physically, it is understood as the quantum-gravitational counterpart of the classical tidal forces $\bar{K}_\mathrm{cl}$ and might therefore give information on the strength of quantum gravitational effects.

\begin{figure*}[t!]
\includegraphics[width=0.51\textwidth]{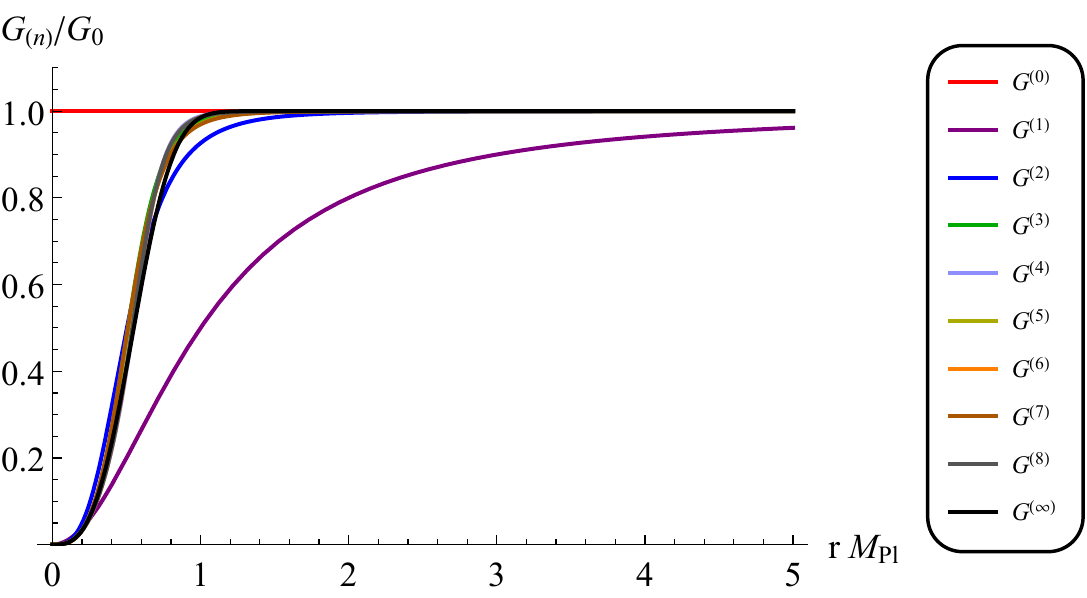} \hspace{0.1cm}\includegraphics[width=0.47\textwidth]{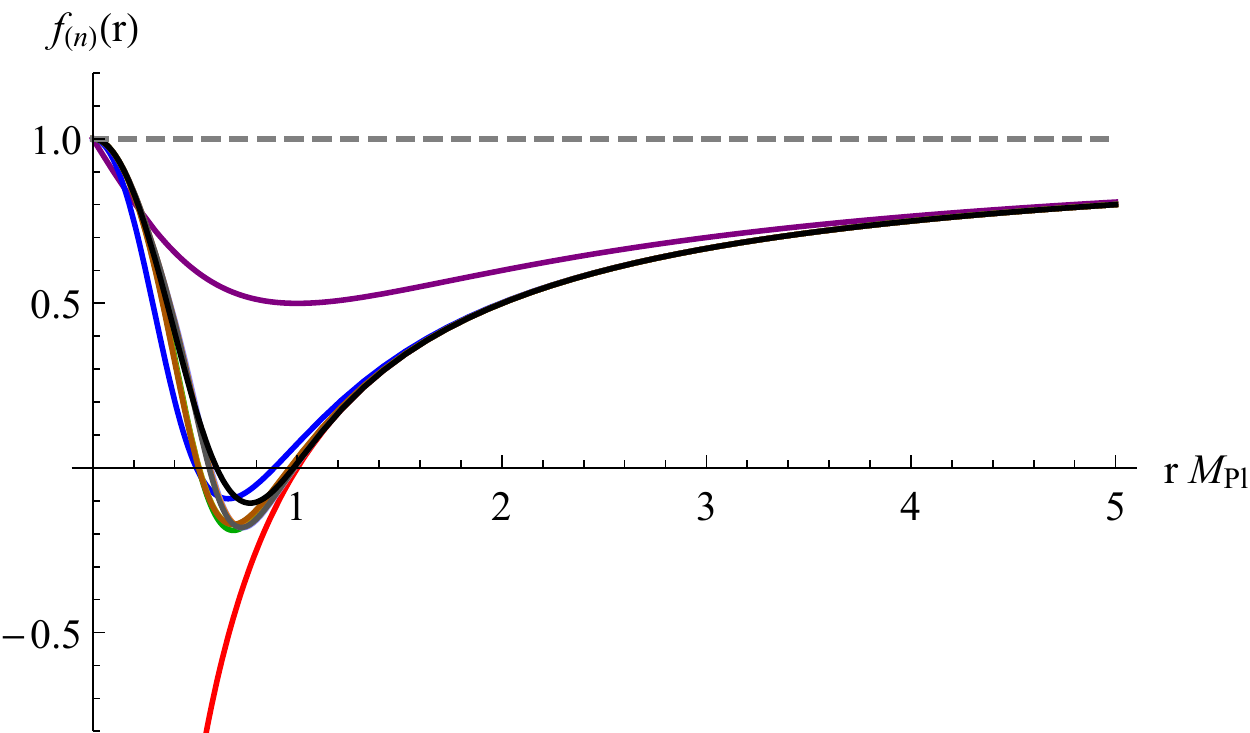}\\[0.3cm]
\includegraphics[width=0.5\textwidth]{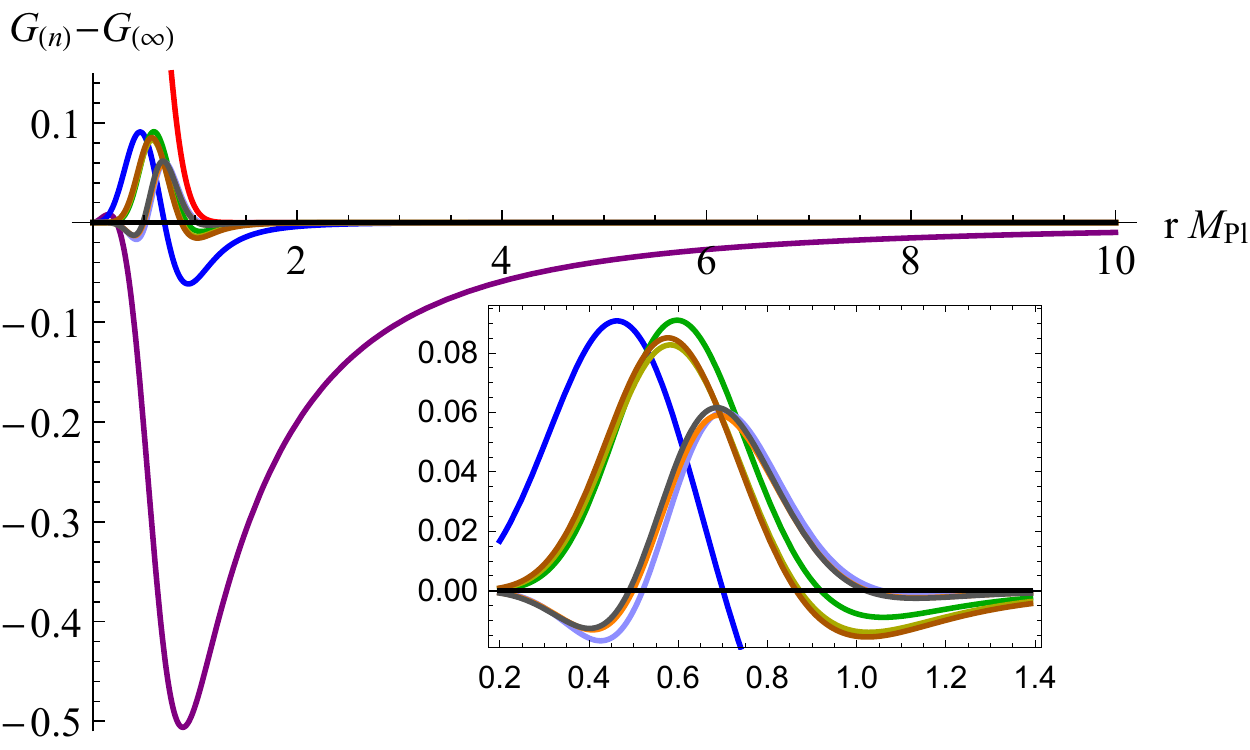} 
\caption{The figures show the sequence of Newton's couplings $G_{(n)}(r)$ (left top panel), $f_{(n)}(r)$-functions (right top panel) defined by the RG-improvement iteration~\eqref{RGiteration}, and its convergence properties (bottom panel). The red lines represents the ``step 0'' of the sequence and correspond to a classical Schwarzschild black hole. The sequence of functions shown in the figure has been obtained by using the initial condition $k_{(1)}(r)=\xi/r$ and the values $m=M_\mathrm{Pl}$ and $\xi\sim g_\ast \sim1$.  In this particular setting, the sequence~\eqref{RGiteration} rapidly converges to the Newton's coupling and $f(r)$-function associated with the Dymnikova solution~\eqref{Ginf} (black lines). A different initial condition changes the convergence properties of the sequence, but does not affect the final solution~\eqref{Ginf}.} \label{Fig}
\end{figure*}

Under these assumptions, the emergent mass scale $(G_0\,\rho)$ can be interpreted as an infrared cutoff-function and we can set
\be\label{eq:cutoffrel}
{k^2}\equiv \mathcal{K}[\rho]=\xi\,G_0\,\rho \;,
\ee
where $\xi$ is a positive dimensionless constant. Replacing the constraint~\eqref{eq:cutoffrel} into eq.~\eqref{fpeq} yields
\be
G_{(\infty)}(r)=\frac{G_0}{1+\frac{\xi g_\ast^{-1}G_0^2m G_{(\infty)}'(r)}{4\pi r^2 G_{(\infty)}(r)}}\;\;.
\ee
This differential equation can be solved analytically and its solution reads
\be \label{Ginf}
G_{(\infty)}(r)=G_0\left\{1-\mathrm{exp}\left(-\frac{r^3}{r_s \,l_{cr}}\right)\right\}\;,
\ee
where $r_s$ is the classical Schwarzschild radius, and the length
\be
l_{cr}=\sqrt{\frac{3\xi}{8\pi g_\ast}} l_\text{Pl}\;,
\ee
represents a critical length-scale below which the modifications induced by the running of the Newton's constant become important. Assuming $\xi\sim g_\ast\sim1$ this critical scale is of the order of the Planck length, $l_\text{Pl}$ \cite{br00,2006alfiomartin}. The first eight steps of the sequence of RG-improvements~\eqref{RGiteration} and its fixed point solution \eqref{Ginf} are shown in Fig. \ref{Fig}.

It is straightforward to see that the resulting metric corresponds to a (regular) black-hole solution of the Dymnikova-type \cite{Dymnikova:1992ux,Dymnikova:1996ux}. Expanding the corresponding lapse function
about $r=0$ yields
\be
f_{(\infty)}(r)\sim 1-\frac{r^2}{l^2_{cr}}+o(r^5)\;\;.
\ee
The running Newton's coupling $G_{(\infty)}(r)$ hence generates a de-Sitter core \cite{br00,Koch:2013owa,Adeifeoba:2018ydh} characterized by the following microscopic induced cosmological constant
\be
\Lambda=\frac{8\pi g_\ast}{\xi \,G_0}\;.
\ee
It is important to notice that the case $g_\ast\to0$, i.e. asymptotically-free gravity, would require the running Newton's constant $G_{(\infty)}(r)$ to be everywhere zero. The case $g_\ast\to\infty$ (no fixed point) would instead imply $G_{(\infty)}(r)\equiv G_0$, and reproduces the classical (singular) Schwarzschild spacetime. The only non-trivial solution, namely a spacetime metric that is everywhere regular and reproduces the classical solution at large distances, is realized when the fixed-point value of the dimensionless Newton's constant is finite and non-zero, $g_\ast\neq0$, i.e. when the gravitational RG flow is asymptotically safe. This finding strengthens the possibility that the perturbative non-renormalizability of gravity and the appearance of singularities in classical General Relativity might indeed be related \cite{Adeifeoba:2018ydh}, and that a proper application of the RG setup might lead to an effective renormalization of the spacetime geometry that removes the typical singularities affecting the classical theory.

\section{Conclusions} \label{sect4}

In this work the effects of a running Newton's coupling on classical Schwarzschild black holes have been investigated using an iterative renormalization group (RG) improvement procedure. This semi-classical approach allows to take into account the backreaction produced by the running of the Newton's coupling on the cutoff function and the Einstein field equations at the same time. 

Imposing the scale-dependence of the gravitational coupling at the level of the classical background geometry causes the appearance of an effective energy-momentum tensor, encoding vacuum polarization effects of the (quantum) gravitational field \cite{Poisson:1988wc,Poisson:1990eh}. This results in a ``quantum tidal force'' whose intensity is proportional to the temporal component of the effective stress-energy tensor, $\rho\propto \partial_r G(r)$.  Relating the RG scale to this effective energy-density leads to a simple recursive relation which describes how the spacetime metric backreacts onto the variation of the Newton's coupling. 

Assuming that the sequence of RG-improvements converges, the limiting fixed-point solution is given by a simple first-order differential equation that can be solved analytically. Due to a cancellation between the classical tidal forces, described by the classical Kretschmann scalar $\bar{K}_\text{cl}$, and their quantum counterpart, controlled by the tension $\rho$, the resulting ``renormalized'' tidal forces are finite and the spacetime is devoid of singularities. This cancellation is only possible under the condition of asymptotic safety, i.e. only if the gravitational coupling approaches an interacting fixed point in the high-energy limit. In this case the sequence of RG-improvements converges to a black-hole spacetime of the Dymnikova-type \cite{Dymnikova:1992ux,Dymnikova:1996ux}. The running of the gravitational coupling, which is assumed to mimic quantum-gravity effects, effectively removes the singularity affecting the classical solution and, in this respect, induces an effective renormalization of the bare spacetime geometry.

Interestingly, the appearance of an effective energy-momentum tensor in the improved Einstein field equations might be interpreted as the result of new non-Einsteinian terms in the effective Lagrangian. These novel terms are due to the effective renormalization process induced by including the running of the Newton's coupling into the classical field equations. Finding an effective action which reproduces the modifications found in the present work would be highly desirable, as it might give hints for the reconstruction of the quantum effective action based on the combination of non-perturbative functional renormalization group technique with results from causal dynamical triangulation \cite{Knorr:2018kog}. Finally, an important question to tackle is the stability of the result under the introduction of a cosmological constant term: when the running of the cosmological constant is considered, the scale-setting $k=k[\rho(r)]$ is entirely dictated by the requirement of preserving diffeomorphism invariance~\cite{guberina05}. This would remove from the present construction all ambiguities related to the scale identification. We hope to address the aforementioned tasks in future works.

\begin{acknowledgments}
The authors thank A. Bonanno, A. Eichhorn and K. Stelle for important comments and discussions. This work was supported by the DFG under grant no Ei-1037/1 and by the Alexander von Humboldt Foundation.
\end{acknowledgments}

\bibliography{AleBib}

\end{document}